\documentclass[article]{jfm}

\usepackage{graphicx}
\usepackage{newtxtext}
\usepackage{newtxmath}
\usepackage{natbib}
\usepackage{hyperref}
\hypersetup{
	colorlinks = true,
	urlcolor   = blue,
	citecolor  = black,
}

\newcommand{\RomanNumeralCaps}[1]

\usepackage{subfigure}
\usepackage{todonotes}

\usepackage{xparse}

\newcommand{\dd}[1]{\mathrm{d}#1}

\newcommand{\matr}[1]{\textbf{#1}} 
\newcommand{\vect}[1]{\boldsymbol{#1}}

\newcommand{\pder}[2]{\frac{\partial#1}{\partial#2}} 
\newcommand{\tder}[2]{\frac{\dd#1}{\dd#2}} 

\title{Advancing Fluid Dynamics Stability Analysis: Construction of Lyapunov Functions via the Generalized Kinetic Energy Approach}

\author{P\'{e}ter Tam\'{a}s Nagy\aff{1}
	\corresp{\email{pnagy@hds.bme.hu}},
}

\affiliation{\aff{1}Department of Hydrodynamic Systems, Faculty of Mechanical Engineering, Budapest University of Technology and Economics. M\H{u}egyetem rkp. 3., H-1111 Budapest, Hungary
}
\begin{document}
	\maketitle
	
	\begin{abstract}
		
		The energy method, also known as the Reynolds-Orr equation, is widely utilized in predicting the unconditional stability threshold of shear flows owing to the zero contribution of nonlinear terms to the time derivative of perturbation kinetic energy. However, it often underestimates the critical Reynolds numbers compared to experimental measurements.  On the other hand, linear stability analysis tends to yield impractically high limits due to the occurrence of subcritical transitions.
		
		A novel methodology is introduced to enhance and validate the generalized kinetic energy formulation, aiming to provide a more accurate estimation of transition.
		This method considers the influence of nonlinear terms in calculating the threshold amplitude.
		The efficacy of this approach is showcased through the utilization of basic low-order turbulence models and the Poiseuille flow as illustrative examples.
		
		Through the proposed technique, the objective is to bridge the disparity between theoretically predicted critical Reynolds numbers and experimental observations, thus providing a more precise evaluation of shear flow stability. This research contributes to the advancement of stability analysis methods, offering practical implications for diverse fluid flow scenarios.

	\end{abstract}
	
	
	
	\section{Introduction}
	\label{sec:introduction}
	%
	Up to a specific Reynolds number, it is widely acknowledged that most fluid dynamic systems are unconditionally stable \citep{Reynolds1895,Orr1907}. However, beyond this threshold, the behavior of the fluid remains an open question.
	In the 19th century, Lord Kelvin \citep{Kelvin1887} suggested that the stability threshold amplitude decreases as viscosity approaches zero:"... the steady motion is stable for any viscosity, however small; and that the practical unsteadiness pointed out by Stokes forty-four years ago and so admirably investigated experimentally five or six years ago by Osbourne Reynolds, is to be explained by limits of stability becoming narrower and narrower the smaller is the viscosity." Unfortunately, determining this permissible perturbation level of the laminar state has proven to be a challenging problem. The only exception is the well-known linear stability limit, beyond which the laminar state's region of attraction vanishes. While calculating this limit is computationally intensive for general geometries, it is feasible. However, for many practical applications, this limit is excessively high, if not infinite.
	
	The initial solutions for the unconditional stability limit of plane Poiseuille flow were derived by \citet{Reynolds1895} and \citet{Orr1907}. They aimed to minimize the Reynolds number at which the kinetic energy of the disturbance does not grow. This optimization (the Euler-Lagrange function) led to a general eigenvalue problem, where the Reynolds number acted as the eigenvalue. Below the critical value, any perturbation decays exponentially. Initially, solutions were obtained for the two-dimensional problem due to its complexity. However, the computed value, approximately $\Rey=88$, based on the Reynolds number defined by the maximum velocity and half the channel gap, was an order of magnitude smaller than the experimentally observed value.
	Later, \citet{Joseph1969} tackled the three-dimensional problem and revealed that the kinetic energy of spanwise oscillating perturbations could grow at a significantly smaller Reynolds number, specifically 49.55. Additionally, they demonstrated that the most unstable perturbations of two-dimensional base flows were those oscillating exclusively in the spanwise direction, instead of in the streamwise one. Lately, another proof of the same statement was published by \citet{xiong2019}.
	
	Recently, \citet{Falsaperla2019} challenged this established understanding, demonstrating that by redefining the energy norm, purely streamwise oscillating waves emerge as the most critical. Their findings were in excellent agreement with experiments conducted by \citet{Prigent2003}. Moreover, their results aligned with the work of \citet{Moffat1990}, who established the stability of flow perturbed by spanwise oscillating waves. 
	This latter statement were verified by numerical experiments \citep{Lundbladh1994, Reddy1998} where the evolution of perturbed flows were simulated numerically. They found that additional noise was needed for the initial perturbation in the case of purely streamwise or spanwise oscillating flows.
	A further generalization of the kinetic energy was recently investigated by \citet{Nagy2023b}, who introduced multiplicators in the definition of kinetic energy for all velocity components. Addressing the three-dimensional domain, they predicted a critical Reynolds number roughly 25\% larger for both Couette and Poiseuille flows. Their analysis indicated that critical perturbations manifest as tilted waves in both flow configurations. However, it's worth noting that their study neglected a non-linear term in pressure calculations, limiting its validity to a specific perturbation amplitude; this limit, however, was not determined. 
	The present research is the continuation of their idea. The definition of kinetic energy is further generalized, and the developed method can predict the threshold amplitude. The definition of this generalized kinetic energy is equivalent to the definition of \citep{Nerli2007}, who redefined the norm by a perturbation and found relatively accurate threshold amplitude in the case of low-dimensional models of shear flows. 
	
	An alternative approach to enhance the Reynolds-Orr method involves the utilization of enstrophy. \citet{Synge1938} explored this method, and more recently, \citet{Fraternale2018} applied it, predicting a significantly larger critical Reynolds number of $\Rey_\mathrm{crit} = 155$ for the two-dimensional case. Notably, this value is approximately double the energy limit for the same configuration.
	Unfortunately, the non-linear term in the vorticity equation cannot be eliminated in the case of three-dimensional flows. Furthermore, \cite{Nagy2022} showed that in the case of three-dimensional systems the predicted critical Reynolds number is smaller than in the case of using the original Reynolds-Orr equation even if the non-linear terms are neglected.
	
	Another way of improving the original energy method involves constraining the potential perturbation field rather than altering the definition itself. Originally, such a constraint was that the velocity field must satisfy the continuity equation, implying divergence-free velocity in the context of incompressible flow. \citet{Nagy2023} observed that the solution of the Reynolds-Orr equation fails to meet the compatibility condition essential for a smooth, physically realistic solution. They introduced this condition as a constraint into the problem; however, their ultimate finding was that while the solution of the Reynolds-Orr equation does not meet the condition, there exist velocity fields close to the solution that do fulfill the compatibility condition. This implies that the condition subtly modifies the original result. 
	Another form of restriction was applied in the receptivity problem of compressible boundary layers by \citet{Kamal2023}. They limited the possible excitation fields to physically relevant cases and achieved excellent agreement with simulation results. However, the drawback of their approach lies in the subjective nature of selecting physically relevant perturbations, which can be highly dependent on the specific flow configuration.

	In the aforementioned cases where stability was established, the non-linear terms of the Navier-Stokes equations were either eliminated or treated as zero. Yet, it is likely that further improvements can only be achieved by considering these terms. One promising approach is to regard the non-linear part as an excitation and establish a bound for it, thus obtaining conditional stability. This concept was explored in the context of Couette flow using the resolvent of the linear operator in the unstable half-plane by \cite{Kreiss1994}. However, extending this solution method further appears to be challenging.
	Another, more comprehensive method that models the non-linear term as a bounded excitation of the linear system has been developed by two groups: \citet{Liu2020} and \citet{Kalur2021}. Referred to as the quadratic constrained (QC) method, they applied this technique to simple turbulence models.
	Alternatively, a broader approach to constructing Lyapunov functions is the sum-of-squares method. In the realm of fluid dynamics, \citet{Goulart2012} proposed the utilization of this technique to establish the global stability of fluid dynamic problems. They demonstrated its effectiveness on a ninth-order model of Couette flow. \citet{Fuentes2022} employed this optimization technique to create non-quadratic Lyapunov functions. They projected the velocity field onto the modes of the classic energy equation solutions and achieved a significantly higher Reynolds number limit using 13 modes.
	While this method holds promise in constructing Lyapunov functions, its computational demands increase rapidly as the number of dimensions grows \citep{Liu2020}.

	Recently, \citet{Pershin2020} introduced a probabilistic approach to assess the stability of Couette flow. Additionally, they proposed a control technique aimed at expanding the region of attraction of the laminar state.

	A fundamentally different approach to address this problem involves calculating the minimal perturbation necessary to induce a non-laminar solution, often referred to as the minimal seed. This approach is similar to conditional stability calculations; however, in this methodology, optimization occurs on the unstable side of the boundary between the stable and unstable regions. Implicitly, the existence and realization of these minimal seeds demonstrate stability, as the flow must remain stable below the perturbation amplitude of the minimal seed.
	
	The first attempts to find such state began in the 1990s. In the initial approaches \citep{Kreiss1994, Lundbladh1994, Reddy1998, Andersson1999}, researchers introduced perturbations that were solutions of linear or energy stability analyses, or they optimized the growth of the linear system. The perturbation amplitude was minimized to establish the threshold level. With advancements in computational capacity, it became possible to optimize the perturbation of the full non-linear problem. Typically, the initial kinetic energy is minimized, leading to maximal kinetic energy after a certain time horizon. For low-order flow models proposed by Waleffe \citep{Waleffe1995, Waleffe1997}, \citet{Cossu2005} calculated the energy of these minimal seeds. Later, this method was applied to real flow configurations \citep{Cossu2005, Duguet2013, Kerswell2014, Kerswell2018, Parente2022, Zhang2023}. Non-linear optimizations revealed localized perturbation fields \citep{Wu2023} with significantly lower kinetic energy than perturbations optimized by linear methods. Readers are referred to the cited papers for a more detailed discussion and specific results.
	Comparing these minimal seed results with threshold amplitude values from stability analyses can be instrumental in estimating the methods' accuracy. If they closely align, it suggests a well-modeled boundary between the stable and unstable regions. However, if they differ significantly, it signals the need for further development in at least one of the methods.

		In this paper, the classic energy method is presented for discretized fluid mechanical systems. Then, the generalized kinetic energy (GKE) method is introduced in Section \ref{sec:theory}. The method is first applied to simple equations of turbulence: the Threfethen two-dimensional TTRD' model \citep{Baggett1997} (Section \ref{sec:TTRDp}) and the Waleffe 1995 (W95) model \citep{Waleffe1995} (Section \ref{sec:W95}). In the next step, the method is demonstrated for higher, yet still relatively low-order models of Poiseuille flow with 180 and 520 degrees of freedom (Section \ref{sec:Poi_flow}). These models are created using the Galerkin projection method, employing the Stokes eigenfunctions.
		
		Finally, the findings and conclusions are summarized in Section \ref{sec:Conclusion}.

		\section{Theory}\label{sec:theory}
		\subsection{The original energy method}
		
		

		When employing Galerkin or Galerkin-Petrov projection on the perturbed Navier-Stokes equation, the perturbed fluid motion can be described by the following ordinary differential equation system \citep{Nerli2006}:
		
		\begin{equation} \label{eq_orig_ODE}
			\tder{q_i}{t} = A_{i,j}\,q_j + Q_{i,j,k}\, q_j\, q_k,
		\end{equation}
		$q_i(t)$ represents an $n$-element vector ($i=1...n$) describing the perturbation of the base flow over time $t$. The coefficients $A_{i,j}$ and $Q_{i,j,k}$ are time-independent arrays characterizing the behavior of the perturbed flow, where $i, j,$ and $k$ are running variables ranging from $1$ to $n$ in the Einstein summation notation. For convenience, the last term in equation (\ref{eq_orig_ODE}) can be rewritten as:
		\begin{equation}\label{eq_non_linear_matrix}
			Q_{i,j,k}\, q_j\, q_k = N_{i,j}(q_i), 
		\end{equation}
		where
		\begin{equation}
			N_{i,j}(q_i) =  Q_{i,j,k}\, q_k.
		\end{equation}
		
		The investigated system is stable, if the perturbations ($q_i$) tend to zero as $t\to\infty$. 
		In cases where the perturbation is assumed to be small ($q_i\propto\epsilon$), neglecting the non-linear (quadratic) terms in the equation allows for linear stability analysis. This involves examining the eigenvalues of the matrix $A_{i,j}$. However, such an analysis is often insufficient in practical applications. 
		$A_{i,j}$ is non-normal meaning that the eigenvectors are non-orthogonal. For small initial perturbations, the amplitudes can grow exceptionally large and the non-linear terms cannot be neglected \citep{Schmid2007, Kerswell2018}. 
		
		An alternative method of stability analysis involves examining the derivative of the perturbation kinetic energy with respect to time. Assuming the kinetic energy of the perturbations is the inner product of the state vector: 
		\begin{equation}\label{eq:def_e}
			e = q_i q_i, 
		\end{equation}
		its temporal derivative can be easily obtained from equation (\ref{eq_orig_ODE}):
		\begin{align}
			\tder{e}{t} &= 2\, q_i \tder{q_i}{t} \\
			\tder{e}{t} &=  2\, A_{i,j}\,q_i\,q_j + 2 \, Q_{i,j,k}\,q_i\, q_j\,q_k . \label{eq_de_dt}
		\end{align}
		According to the Reynolds-Orr identity \citep{Orr1907, Schmid2001} (utilizing Gauss divergence theorem), the non-linear term does not influence the change in kinetic energy if the perturbations are confined by walls, are periodic, or decay to zero in the far field in directions, which are reasonable assumptions in most cases.
		\begin{equation}\label{eq_non_lin_zero_energy_change}
			2\,Q_{i,j,k}\,q_i\, q_j\,q_k = 0.
		\end{equation}
		From this point, matrices and vectors are denoted by bold letters to enhance readability. The Einstein summation notation is used when a three-dimensional array appears in an expression or the discussion.

		\textit{
			It is important to note that if the product 
			$\vect{q}^T \vect{q}$ is not equal to the kinetic energy, the Reynolds-Orr identity cannot be applied, and the non-linear term cannot be eliminated. Let's consider an ordinary differential equation system given where the variable is 
			$\tilde{\vect{q}}$ and the kinetic energy can still be calculated as:
			\begin{equation}
				e =\tilde{\vect{q}}^T \matr{W} \tilde{\vect{q}}  
			\end{equation}
			where $W_{i,j}$ is a real, positive definite matrix typically expressing integration weights. $\matr{W} = \matr{F}^T \matr{F} $ can be obtained using Cholesky decomposition on $\matr{W}$. Since
			\begin{equation}
				e = \tilde{\vect{q}}^T \matr{F}^T \matr{F} \tilde{\vect{q}},
			\end{equation}
			by defining $\vect{q} = \matr{F} \tilde{\vect{q}}$,  $\vect{q}^T\vect{q}$ represents  the kinetic energy. Through the transformation, a new $A_{i,j}$ matrix and $Q_{i,j,k}$ array can be obtained, enabling the application of the Reynolds-Orr identity to the quadratic term.
		}

		The growth rate of the kinetic energy is
		\begin{equation}\label{eq_mu_e_def}
			\mu_e = \frac{1}{e}\tder{e}{t}
		\end{equation}
		and using equations (\ref{eq_de_dt}) and (\ref{eq_non_lin_zero_energy_change}) the following expression can be derived:
		\begin{equation}\label{eq_mu_e_calc}
			\mu_e = \frac{2\, \vect{q}^T \matr{A} \vect{q}}{\vect{q}^T \vect{q}}  
		\end{equation}
		The flow  is considered Lyapunov stable, if $\mu_e<0$ for any $q_i$ state.
		%
		This statement is equivalent to ensuring that the maximum over any possible state is negative:
		\begin{equation}\label{eq_stability_cond}
			\mu_{\mathrm{m},e}=	\max_{\vect{q}} \mu_e(\vect{q})<0.
		\end{equation} 
		%
		%
		The numerator in (\ref{eq_mu_e_calc}) can be written as $2 \vect{q}^T \matr{A}\vect{q} = \vect{q}^T (\matr{A} +\matr{A}^T)\vect{q}$. Moreover, the expression (\ref{eq_mu_e_calc}) represents the Rayleigh quotient of $\matr{A} +\matr{A}^T$. Since $\matr{A} +\matr{A}^T$ is a symmetric matrix, the largest Rayleigh quotient corresponds to the largest eigenvalue of $\matr{A} +\matr{A}^T$, which is the maximum possible growth rate of kinetic energy.
		Therefore, the flow is Lyapunov stable if the largest eigenvalue of $\matr{A} +\matr{A}^T$ is negative:
		
		\begin{equation}\label{eq_mu_max_eig_val}
			\lambda_{\max}\left(\matr{A} +\matr{A}^T\right) <0.
		\end{equation}
		
		The critical state, which maximizes the growth rate of kinetic energy, is the corresponding eigenvector.
		Unfortunately, this condition is strict for practical application. This analysis is referred to as energy method or non-linear stability analysis since the results are valid for the non-linear system due to the non-linear terms not being assumed zero during the derivation but were eliminated by the Reynolds-Orr identity.

		In many fluid dynamic applications, the concern is not just whether the flow is stable or not, but what the limit is where the flow becomes unstable. It's important to note that viscosity or the Reynolds number only affects a specific part of the linear terms ($\matr{A}$) because the Laplace operator in the Navier-Stokes equation is linear and does not directly influence the non-linear terms. Let us decompose the matrix $\matr{A}$ into components dependent on the Reynolds number and those independent of it:
		\begin{equation}\label{eq_A_decomp}
			\matr{A}(\Rey) = \matr{A}_U + \frac{1}{\Rey} \matr{A}_R.
		\end{equation}
		
		Considering that the Laplacian term can only dissipate kinetic energy, 
		$\matr{A}_R$ is a positive definite matrix. The smallest Reynolds number, where $\mu_{\mathrm{m},e} = 0$  , is equivalent to the smallest Reynolds number where $\mu_e=0$. By substituting (\ref{eq_A_decomp}) into (\ref{eq_mu_e_calc}), setting the expression to zero, and subsequently expressing $\Rey$ and calculating its minimum through variation, we arrive at the corresponding Euler-Lagrange equation: 
		\begin{equation}\label{eq_Re_E}
			\matr{A}_R + \matr{A}^T_R = \tilde\Rey\left(-\matr{A}_U -\matr{A}^T_U\right).
		\end{equation}
		This equation represents a general eigenvalue problem where the eigenvalue is the Reynolds number. The smallest eigenvalue, typically denoted as $\Rey_\mathrm{E}$, is referred to as the global stability limit. If $\Rey<\Rey_\mathrm{E}$, then $\mu_{\mathrm{m},e}<0$, signifying unconditional stability in the flow.

		\subsection{The generalized energy method}
		The classical energy method often proves to be highly conservative, predicting Reynolds number limits below experimental observations. This issue arises because at high Reynolds numbers, the $\matr{A}$ matrix becomes non-normal. In such cases, the eigenvectors are non-orthogonal, and even in a linearly stable system, energy can grow significantly \citep{Schmid2007} although it does not necessarily lead to a turbulent state.
		
		The key to improving this method lies in introducing a generalized kinetic energy formulation, a concept also proposed by \citet{Nerli2007}. The transformation of the state vector $\vect{q}$ by an invertible $\matr{S}$ matrix is given by 
		\begin{equation}
			\vect{q} =\matr{S}\,\vect{r}, 
		\end{equation}
		and the generalized kinetic energy is defined as
		\begin{equation}\label{eq:def_gen_kin_en}
			h= \vect{r}^T\vect{r}.	
		\end{equation}
		This definition of generalized kinetic energy is equivalent to the one proposed by \citet{Nerli2007}. However, their approach involved redefining the norm using a perturbation matrix, while here, variables are transformed. Although the objective of determining the allowable perturbation level is similar, the construction of the new norm is different. Additionally, the solution technique for calculating the threshold amplitude (defined in equation (2.3) in \citet{Nerli2007}) was not detailed there, a critical aspect for large systems.

		The differential equation (\ref{eq_orig_ODE}) can be rewritten as:
		\begin{equation}\label{eq_new_ODE_pre}
			\tder{\,S_{i,j} r_j}{t} = A_{i,j}\,S_{j,k}\,r_k + Q_{i,j,k}\, S_{j,l}\, r_l \, S_{k,m}\,r_m
		\end{equation}
		and 
		\begin{equation}\label{eq_new_ODE}
			\tder{ r_i}{t} = S_{i,j}^{-1} A_{j,k}\,S_{k,l}\,r_l + S_{i,j}^{-1} Q_{j,k,l}\, S_{k,m}\, r_m \, S_{l,o}\,r_o.
		\end{equation}
		
		To facilitate this transformation, let's define:
		\begin{align}%
			\tilde{A}_{i,j} &= S_{i,l}^{-1} A_{l,k}\,S_{k,j}\label{eq_A_tilde_def}  \\
			\tilde{Q}_{i,j, k} &= S_{i,m}^{-1} Q_{m,o,l}\, S_{o,j}\, S_{l,k} \label{eq_Q_trans} \\
			\tilde{N}_{i,j}(r_i) &= S_{i,o}^{-1} Q_{o,k,l}\, S_{k,m}\, r_m \, S_{l,j} 
		\end{align}
		These transformations result in a similar ordinary differential equation as (\ref{eq_orig_ODE}), if ${A}_{i,j}$ and ${Q}_{i,j, k}$ are replaced by $\tilde{A}_{i,j}$ and $\tilde{Q}_{i,j, k}$, respectively. It is worth noting that while transforming the coefficient array $Q$ might not be beneficial in practice due to computational expenses, the transformation of state vectors is a more computationally efficient alternative.
		
		The growth rate of the generalized kinetic energy is defined as: 
		\begin{equation}\label{eq_mu_h_def}
			\mu_h = \frac{1}{h}\tder{h}{t},
		\end{equation}
		and can be calculated similarly to (\ref{eq_de_dt}) as
		\begin{equation}\label{eq_mu_h_calc}
			\mu_h = \frac{2\, \tilde{A}_{i,j}\,r_i\,r_j + 2 \, \tilde{Q}_{i,j,k}\,r_i\, r_j\,r_k}{r_l r_l}.
		\end{equation}
		The flow is stable, if the $\mu_h<0$ for any state $r_i$. 
		
		The main difference lies in the quadratic term ($\tilde{Q}_{i,j, k}$)  contributing to the growth rate of generalized kinetic energy (\ref{eq_mu_h_def}), unlike in the case of the original kinetic energy. 

		Due to the presence of this term, conditional stability can be established, and it can be utilized to calculate the threshold amplitude.
		
		It is convenient to rewrite the state vector as the product of its magnitude $\gamma = \sqrt{r_i r_i}$ and a unitary vector:
		\begin{equation}\label{eq_def_unitary_r}
			r_i = \gamma \tilde{r}_i.
		\end{equation}
		After substitution into equation (\ref{eq_mu_h_calc}), the growth rate of the generalized kinetic energy can be expressed as:
		\begin{equation}\label{eq_mu_h_calc2}
			\mu_h = 2\, \tilde{A}_{i,j}\,\tilde{r}_i\,\tilde{r}_j + 2\, \gamma\, \tilde{Q}_{i,j,k}\,\tilde{r}_i\, \tilde{r}_j\,\tilde{r}_k.
		\end{equation}
		This approach was also employed by \cite{Nerli2007}. 
		The value $\gamma$ can be used to characterize the amplitude of the perturbation.
		Let us define the possible maximum growth rate at a given level of perturbation as:
		\begin{equation}\label{eq_mu_h_max_def}
			\mu_{\mathrm{max},h} (\matr{S},\gamma) = \max_{\tilde{\vect{r}}} \mu_h(\tilde{\vect{r}}, \matr{S},\gamma).
		\end{equation}
		If the growth rate of generalized energy remains smaller than zero up to a certain amplitude ($\mu_h<0$ if $\gamma< \gamma_\mathrm{crit}$), the investigated system is conditionally stable \citep{Bedrossian2017}, and $h$ is a Lyapunov function. Since $\mu_h\leq\mu_{\mathrm{max},h}$, the flow is stable, if $\mu_{\mathrm{max},h}$ is smaller than zero. 
		
\begin{figure}\centering
			\includegraphics[scale=0.75]{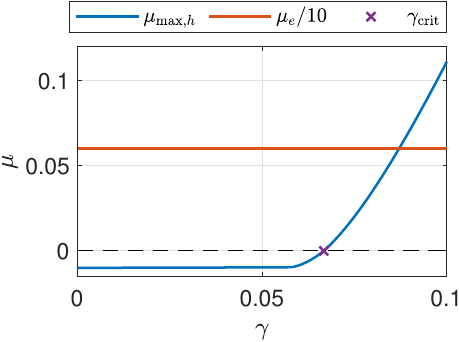}
			\caption{The maximal growth rate of the generalized kinetic energy ($\mu_{\mathrm{max}, h}$) as the function perturbation magnitude $\gamma$ in the case of the optimally transformed TTRD' model at $\Rey = 5$. The red curve represents the one tenth of the growth rate of original kinetic energy ($\mu_e = 0.6$), which is independent of the perturbation level.}
			\label{fig:TTRDp_mu_h_gamma}
		\end{figure}

		The crucial question is how to determine $\gamma_\mathrm{crit}$.  Firstly, it's essential to emphasize that the developed method is applicable to subcritical systems within the investigated range; they must be linearly stable. For a linearly unstable system, $\mu_{\mathrm{max}, h}>0$ for any \matr{S}. In the case of a linearly stable system, there exist transformation matrices where the generalized energy growth rate ($\mu_h$), at least for infinitesimally small perturbations $\gamma \to \epsilon$, practically $\gamma =0$.  As the amplitude of the perturbation ($\gamma$) increases, it can be assumed that the possible maximum growth rate increases continuously. At a certain value, the possible maximum growth rate becomes zero. This value of $\gamma$ is the critical value. It is implicitly defined as:
		\begin{equation}\label{eq_gamma_crit_def}
			\mu_{\mathrm{max},h} (\matr{S},\gamma_\mathrm{crit}) = 0.
		\end{equation}
		The corresponding unitary state vector, defined by 
		\begin{equation}\label{eq_def_r_crit}
			\mu_{\mathrm{max},h} (\matr{S},\gamma_\mathrm{crit}) = \mu_h(\tilde{\vect{r}}_{\mathrm{crit}}, \matr{S},\gamma_\mathrm{crit}),
		\end{equation} 
		can be utilized to obtain the critical state:
		$\vect{r}_{\mathrm{crit}} = \gamma_\mathrm{crit} \tilde{\vect{r}}_{\mathrm{crit}}$. The maximal growth rate of generalized energy ($\mu_{\mathrm{max},h}$) as the function of excitation magnitude $\gamma$ is plotted in figure \ref{fig:TTRDp_mu_h_gamma}. At low $\gamma$ values, the linear part of the dynamical system dominates, where the maximum growth rate is almost constant and equal to the Rayleigh coefficient of the $\tilde{A}_{i,j}+\tilde{A}_{j,i}$ matrix. For higher $\gamma$ values, the non-linearity of the system influences the maximal growth, which tends towards a straight line. The slope of this line corresponds to the maximum of $\left\lbrace  2\, \, \tilde{Q}_{i,j,k}\,\tilde{r}_i\, \tilde{r}_j\,\tilde{r}_k\right\rbrace$ among possible $\tilde{r}_i$ states.

		The investigated region can be envisioned as a multidimensional hypersphere in the 
		$\vect{r}$
		state space around the origin. The radius of this sphere is $\gamma$.  If the radius is smaller than a critical value $\gamma_\mathrm{crit}$, then $\mu_h<0$, indicating that the norm of the solution vectors is decreasing, and the trajectories move inward the sphere, ultimately converging to the origin. At the critical radius, a trajectory becomes tangential to the sphere, and it may not reach the origin. The hypersphere with radius $\gamma_\mathrm{crit}$  represents the stability region. Outside this sphere, the system can be, but is not necessarily, unstable. In the case of the two-dimensional problem, the stability region reduces to a circle and will be illustrated in Subsection \ref{sec:TTRDp} in figure \ref{fig:TTRDp_ROA_r}.
		
		The presented method offers the flexibility of varying and optimizing the transformation matrix. A common approach might be to maximize the stability region described by the value of $\gamma_\mathrm{crit}$ in the state space of 
		$\vect{r}$ vectors. However, this optimization strategy is not advantageous, as multiplying 
		$\matr{S}$  by an arbitrary constant greater than one would inflate $\gamma_\mathrm{crit}$. To address this issue, one option is to constrain the norm of the transformation matrix. However, a more beneficial and informative approach is to transform the stability region back to the original state space of 
		$\vect{q}$. 
		
		The linear transformation (scaling and rotating) of the hyperspehere results in a hyperelipsoid in the original state space 
		$\vect{q}$. This hyperellipsoid defines the boundary of the region of attraction of the origin. Although the kinetic energy ($e$) can grow significantly inside this region, stability is guaranteed due to the exponential decay of the solution in a properly chosen solution norm ($h$). The largest radius of a hypersphere contained within the hyperellipsoid is equal to the smallest minor axis of the hyperellipsoid. The square of this radius ($e_{\mathrm{min}}$) represents the threshold kinetic energy below which the flow remains stable.
		
		The region of attraction in both the original and transformed state spaces is illustrated in figure \ref{fig:TTRDp_ROA} in the case of a two-dimensional turbulence model. Due to the similarities to the method of \citep{Nerli2007}, who utilizes generalized kinetic energy, the region of attraction was a hyperellipsoid, there.
		
		In addition, it is crucial to note that in this context, "$\mathrm{min}$" pertains to the minimum squared radius of the region of attraction, not the minimal energy threshold leading to a turbulent state. The value $e_{\mathrm{min}}$ can be mathematically expressed using equations (\ref{eq:def_e}) and (\ref{eq_def_unitary_r}) as follows:
		\begin{equation}
			e_{\mathrm{min}}(\matr{S}) = \gamma_\mathrm{crit}^2 (\matr{S}) \min_{\tilde{\vect{r}}} \lbrace \tilde{\vect{r}}^T \matr{S}^T \matr{S} \tilde{\vect{r}} \rbrace. 
		\end{equation}
		The argument of the minimum function is the Rayleigh coefficient of $\matr{S}^T \matr{S}$, and the minimum value corresponds to the smallest eigenvalue of $\matr{S}^T \matr{S}$, since $\matr{S}^T \matr{S}$ is a symmetric matrix. 
		\begin{equation}\label{eq:e_min_calc}
			e_{\mathrm{min}}(\matr{S}) = \gamma_\mathrm{crit}^2 (\matr{S}) \;\lambda_\mathrm{min} \left( {\matr{S}^T \matr{S}}\right) .
		\end{equation}
		The corresponding unitary eigenvector 
		$\tilde{\vect{r}}_{\mathrm{min}}$ can be utilized to get the two locations 
		$\vect{q}_{\mathrm{min}} = \pm\gamma_\mathrm{crit}\,\matr{S}\, \tilde{\vect{r}}_{\mathrm{min}} = \pm\matr{S}\, \vect{r}_{\mathrm{min}}$, where the hypersphere touches the hyperelipsoid, , as illustrated in figure \ref{fig:TTRDp_ROA_q}.
		
		If the kinetic energy of the perturbation is smaller than this critical value ($e<e_{\mathrm{min}}$), the flow is stable. The aim of the method is to maximize this limit, $e_{\mathrm{min}}$.
		%
		It is important to note that maximizing the norm of  
		$\vect{q}_{\mathrm{crit}} = \matr{S}\, \vect{r}_{\mathrm{crit}}$ would be unfeasible. Such an optimization would result in singular transformation matrices, causing stability regions to resemble "nail"-like structures.
		
		
		\subsection{The usage of the generalized energy method}
		Two key questions remain unanswered. The first one concerns how the maximal growth rate (\ref{eq_mu_h_max_def}) can be calculated, as it has a non-linear dependence on the state vector.
		Proving that a specific $\tilde{\vect{r}}$ maximizes the expression (\ref{eq_mu_h_calc2}) while satisfying the constraint of unity for the state vectors ($\tilde{\vect{r}}$) is a challenging task. This can be accomplished using Sum of Squares (SOS) methods, although they are computationally very expensive, as highlighted by \citep{Fuentes2022}. Simultaneously, other general constrained optimization techniques have seen significant advancements in recent decades. Typically, these methods compute the minimum rather than the maximum; hence, the functions to be maximized are multiplied by minus one.
		
		It is worth noting that the expression in (\ref{eq_mu_h_calc2}) is analytical, allowing for the analytical and explicit derivation of the gradient and the Hessian matrix. This feature enhances the efficiency of the optimization process. Various methods, including Sequential Quadratic Programming (SQP), Active Set Algorithm, and Interior Point Algorithm \citep{Nocedal2006}, were explored. These methods are implemented in MATLAB's \textit{fmincon} function. After considering factors such as calculation time, accuracy, and robustness of the methods, it was found that the SQP method proved to be optimal for small systems ($n=4$), while the Interior Point Algorithm performed best for larger systems ($n>180$).
		
		During the optimization, multiple random seed vectors were generated to initialize the process. Interestingly, at low $\gamma$ values (less than 0.1), 60-100\% of the cases converged to the same maximum. Even at high perturbation magnitudes ($\gamma\approx10$), the convergence rate remained above 40\%, as observed in the case of the four-dimensional turbulence model by \cite{Waleffe1995}. This observation suggests that the optimization procedure successfully identifies the global maximum.
		
		The second key question is how the optimal transformation matrix $\matr{S}_\mathrm{opt}$ can be obtained.
		
		One plausible approach involves considering the eigenvectors of $\matr{A}$ as the initial choice for $\matr{S}$. 
		They diagonalize the linear part of the system. 
		Under this transformation, the new state variables correspond to the coefficients of the eigenmodes. 
		The generalized kinetic energy is represented as the sum of these coefficient squares, ensuring that the system achieves energetic stability at low perturbation level. 
		Such a transformation solves the issue of non-normality, since the eigenvectors of the transformed system are orthogonal.
		However, it is worth noting that in certain scenarios, $\matr{A}$ might not be diagonalizable. 
		This occurs when the eigenvectors are not linearly independent, rendering the inverse of the transformation matrix non-existent.
		
		Moreover, empirical attempts have revealed that this approach is suboptimal since it fails to maximize $e_\mathrm{min}$, a critical criterion in the optimization process.

		A potential approach for optimizing $\matr{S}$, can be outlined as follows:
		\begin{enumerate}
			\item Solve equation (\ref{eq_gamma_crit_def}) for $\gamma_\mathrm{crit}$.
			\item Calculate $e_\mathrm{min}$ utilizing equation (\ref{eq:e_min_calc}).
			\item Update ($\matr{S}$)  systematically and repeat steps 1 and 2 iteratively until $e_\mathrm{min}$ (\ref{eq:e_min_calc}) converges to its maximum. 
		\end{enumerate}
		%
		This systematic process ensures a step-by-step refinement of $\matr{S}$, allowing the optimization to progress toward the maximum value of $e_\mathrm{min}$.
		
		This method is indeed feasible; however, the absence of gradients poses a significant challenge, especially when dealing with a large number of unknowns ($n^2$), leading to computationally expensive optimizations. While one potential approach involves implicit differentiation of the expression $e_\mathrm{min}(S_{i,j})$, this method proves exceptionally challenging. Implicit differentiation necessitates solving a complex nonlinear equation system, contrasting with the straightforward calculation of an explicit expression. Consequently, in cases where the system comprises a limited number of degrees of freedom, optimization without analytical gradients remains possible. Nonetheless, for expansive systems, the absence of these gradients renders the optimization process unfeasible due to its computational intensity.

		An alternative approach involves introducing $\gamma$ as an additional optimization variable within the elements of the transformation matrix ($\matr{S}$). Simultaneously, a constraint is imposed, mandating the growth rate to be zero. The expression $e_\mathrm{min}(\matr{S} \gamma_\mathrm{crit})$ is optimized, which is constrained by the equation (\ref{eq_gamma_crit_def}). Although this method slightly increases the number of unknowns, it significantly enhances the efficiency of the optimization process. The reason lies in the explicit and efficient calculation of gradients, which become feasible due to this approach.
		
		However, the previously mentioned numerical methods (SQP, Active Set, Interior Point) were not robust enough to handle this problem, likely due to its high sensitivity to the constraint. Initially, an attempt was made using the augmented Lagrangian method \citep{Nocedal2006} where another penalty term is added to mimic the Lagrange multiplier. This multiplier should be updated at each iteration to fulfill the constraint. However, solving the constraint equation (\ref{eq_gamma_crit_def}) for $\gamma_\mathrm{crit}$ in each iteration significantly reduced the computational time significantly due to faster convergence and required fewer iteration step. Therefore, the usage of the Lagrange multiplier term lost its sense and it was abandoned, and the method simplified to the penalty method \citep{Nocedal2006}. The resulting optimization problem in each iteration step is solved by the \textit{fminunc} function using 'quasi-newton' method, and then equation (\ref{eq_gamma_crit_def}) is solved for $\gamma_\mathrm{crit}$ to fulfill the constraint. This modified approach proved to be more effective and computationally efficient.
		Additional essential details regarding the optimization process, including gradients and Hessian matrices of the functions, can be found in Appendix \ref{sec:app_numerical_methods}.
		
		

		\section{Application}
		
		\begin{figure}
			\centering
			\subfigure[]{
				\includegraphics[scale=0.75]{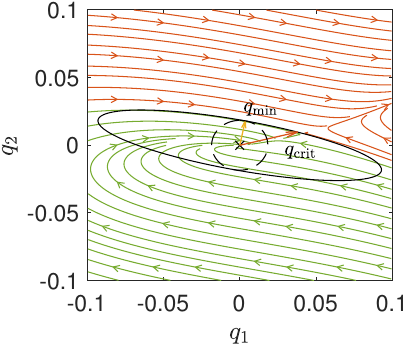}
				\label{fig:TTRDp_ROA_q}} 
			\hspace{5mm}
			\subfigure[]{
				\includegraphics[scale=0.75]{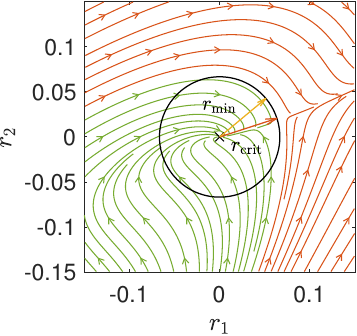}
				\label{fig:TTRDp_ROA_r}} 
			\caption{The phase space trajectories of the TTRD' model at $\Rey = 5$. Green trajectories  converge towards the origin, while red trajectories tend to another equilibrium point, which is not shown. Black curve represents the boundary of the region of attraction. The red vector is the critical perturbation (\ref{eq_gamma_crit_def}), where the growth rate of the generalized kinetic energy was zero at the critical perturbation level (\ref{eq_def_r_crit}). The yellow vector illustrates the smallest perturbation (\ref{eq:e_min_calc}) in the original state space (a) whose length is equal to the critical perturbation in the optimally transformed state space (b). }
			\label{fig:TTRDp_ROA}
		\end{figure}
		
		\subsection{Trefethen's simple model}
		\label{sec:TTRDp}
		One of the simplest low-order representation of turbulent flows is the TTRD' model described by \cite{Baggett1997}. In this model, the linearized part (\matr{A}) becomes non-normal as the Reynolds number increases. Meanwhile, the non-linear part does not affect the growth rate of kinetic energy, as the corresponding matrix remains asymmetric.

		The TTRD' model is represented by the following equation:
		\begin{equation}\label{eq_TTRDp}
			\frac{\mathrm{d}\vec{q}}{\mathrm{d}t}
			=
			\begin{bmatrix}
				-\frac{1}{\Rey} & 1   \\
				0        &-\frac{1}{\Rey} \\
			\end{bmatrix}
			\vec{q}+
			\begin{bmatrix}
				0 & -q_1 \\
				q_1 &0
			\end{bmatrix}\vec{q}.
		\end{equation}
		In the original reference, the state variables are denoted as $\vec{q} = [u, v]^T$. The non-linear part of equation (\ref{eq_TTRDp}) can be expressed as a non-linear array $Q_{i,j,k}$ (\ref{eq_orig_ODE}). The non-zero elements are
		\begin{align}\label{eq_TTRDp_Q}
			Q_{1,1,2} &= -1\\
			Q_{2,1,1} &= 1.
		\end{align}
		The model remains linearly stable for arbitrarily large Reynolds numbers since the eigenvalues (-1/\Rey) remain negative. However, as the Reynolds number increases, the eigenvectors become non-orthogonal. The unconditional stability limit is $\Rey_E = 2$, as determined by equation (\ref{eq_Re_E}). Above this limit, the kinetic energy of the perturbation can grow, but it does not necessarily lead to a turbulent state. 
		
		The generalized kinetic energy method (GKE) is applied to the problem. The optimal transformation matrices $\matr{S}$, that maximize $e_\mathrm{min}$, are calculated at the Reynolds number between 5 and 100 with the step size 2.5.
		The method is demonstrated in Figure \ref{fig:TTRDp_ROA} at $\Rey=5$, where stable trajectories are depicted in green and unstable trajectories in red. These trajectories are plotted as functions of the original variables (Figure \ref{fig:TTRDp_ROA_q}) and the transformed variables (Figure \ref{fig:TTRDp_ROA_r}). The calculated region of attraction appears as an ellipse in the original state space, precisely touching the unstable trajectories. Outside of this region, there are states from which the solution tends toward another equilibrium point and does not return to the origin. 
		
\begin{figure}\centering
			\includegraphics[scale=0.75]{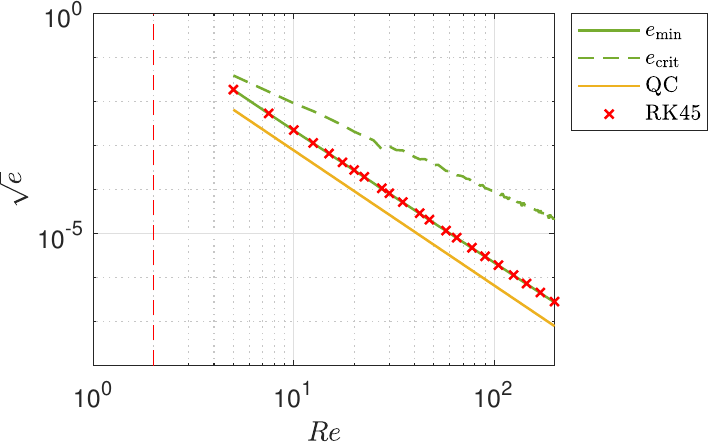}
			\caption{The square root of the smallest kinetic energy at the boundary of the region of attraction ($e_\mathrm{min}$) and the square root of the kinetic energy of the critical perturbation ($e_\mathrm{crit}$) in the case of TTRD' model. The fitted curve from \cite{Liu2020} using the QC method $(0.912\,\Rey^{-3.07})$ is shown alongside. The best-fitting curve of GKE results is $\sqrt{e_\mathrm{min}}\approx 2.228 \,\Rey^{-3.005}$. The red crosses represent the square root of the initial kinetic energy of unstable solutions close to the region of attraction. The vertical red line signifies the unconditional stability limit $\Rey_E = 2$. }
			\label{fig:TTRDp_crit_energy}
		\end{figure}

		The calculated threshold amplitude as the function of Reynolds number are plotted in figure \ref{fig:TTRDp_crit_energy} and compared with the findings of \citet{Liu2020}. The cited authors used the quadratic constraint method, which has been proven to be computationally efficient. They treated the non-linear term as a forcing with an approximated upper bound. The  threshold amplitude is approximated by a power function which is plotted in figure \ref{fig:TTRDp_crit_energy}.

		The calculated threshold amplitude ( $\sqrt{e_{\mathrm{min}}}$) decays as a function of the Reynolds number following a power law in CKE case as well. The exponents are nearly identical: -3.005 in this study and -3.07 in the work of \citet{Liu2020}. However, the method presented here predicts a stable region with a radius roughly three times larger, indicating a energy level approximately one magnitude higher. This substantial difference arises from their approximation of the non-linear term, while GKE calculation takes into account the exact terms, providing a more precise representation of the system's behavior.
		
		In the next step, the accuracy of the region of attraction is investigated by solving the ordinary differential equation close to the outside of the stable region. The solutions are initialized from slightly increased threshold state vectors $\vect{q}_{0} = c_{u} \vect{q}_{\mathrm{min}}$ and computed using the Matlab \textit{ode45} Runge-Kutta method. The $c_{u}$ value is systematically  increased by 0.5\% from 0.98 until an unstable solution obtained. The average value of $c_{u}$ for the unstable solutions is found to be 1.02, indicating that the proposed method is highly accurate; unstable solutions can be obtained very close to the region of attraction. Additionally, the method is partially verified by the observation that in the investigated cases, none of the multipliers fall below one. 
		
		In figure \ref{fig:TTRDp_crit_energy}, the square root of the energy of the critical perturbation ($e_{\mathrm{crit}} = ||\vect{q}_\mathrm{crit}||$) is also plotted. While these values have limited physical relevance in the current study, as they correspond to a critical state in an optimized state space, they can be significant for understanding and analyzing the boundary between laminar and turbulent regions and they could prove useful for further comparisons.
		
		However, it's worth noting that in most cases, these curves show high sensitivity to the optimization convergence, indicating that the results are likely less accurate compared to the $e_{\mathrm{min}}$ values.

		The optimal transformation matrix is 
		\begin{equation}\label{eq_Sopt_TTRDp_Re5}
			\matr{S}_\mathrm{opt} \approx
			\begin{bmatrix}
				0.932426 &  -1.03515\\
				0.0273741 &  0.390105
			\end{bmatrix}
		\end{equation}  
		at $\Rey=5$. 

		\subsection{Waleffe model}
		\label{sec:W95}
		In the next step, the GKE method is applied to the low-order turbulence model proposed by \cite{Waleffe1995}. Since the method under consideration is capable of investigating systems around the origin of the state space, and the laminar equilibrium point in the original model was non-zero, the last state variable was shifted as $n=m-1$ (using the original notation). This adjustment was made following the approach of \citet{Henningson1996} and \citet{Kalur2021}. Consequently, the resulting dynamical system is represented as follows: 
		\begin{equation}\label{eq_W95}
			\frac{\mathrm{d}\vec{q}}{\mathrm{d}t}
			=\frac{1}{\Rey}
			\begin{bmatrix}
				-\lambda_w & \Rey   &  0  & 0\\
				0        &-\mu_w &  0  & 0\\
				0        & 0   &-\nu_w &  0 \\
				0        &  0  & 0   &-\sigma_w \\
			\end{bmatrix}
			\vec{q}+
			\begin{bmatrix}
				-\gamma_w q_3^2 + q_2 q_4 \\
				\delta_w  q_3^2     \\
				\gamma_w q_3 q_4-\delta_w q_3 q_4        \\
				-q_4 q_2        \\
			\end{bmatrix}.
		\end{equation}
		The parameters $\lambda_w, \mu_w, \nu_w, \sigma_w$ represent the decay rates due to viscosity, while $\gamma_w, \delta_w$ describe the non-linear interaction between rolls ($q_2$) and streaks ($q_1$). For a more comprehensive physical explanation of the model, readers are referred to the original paper by \cite{Waleffe1995}.

		The non-linear part of the equation (\ref{eq_W95}) can also be be expressed as:
		\begin{equation}
			\matr{N} = 
			\begin{bmatrix}
				0        & 0 		     &  -\gamma_w q_3 & q_2 \\
				0        & 0		     &  \delta_w q_3  & 0 \\
				\gamma_w q_3 & -\delta_w q_3 & 0            & 0 \\
				-q_2       &  0            & 0            & 0 \\
			\end{bmatrix}
		\end{equation}
		or using the three-dimensional array $Q_{i,j,k}$, where the non-zero elements are:
		\begin{align}
			Q_{1, 2, 4} &= 1;\;\;
			&Q_{1, 3, 3} &= -\gamma_w;\\
			Q_{2, 3, 3} &= \delta_w;\;\;
			&Q_{3, 3, 1} &= \gamma_w;\\
			Q_{3, 3, 2} &= -\delta_w;\;\;
			&Q_{4, 2, 1} &= -1;
		\end{align}

		In this study, three different parameter sets are investigated. The first set is characterized by $\lambda_w = \mu_w= \sigma_w = 10,\, \nu_w = 15,\, \delta_w = 1,\, \gamma_w = 0.1$, denoted as the W95A model \citep{Waleffe1995}. 
		The parameters of the second set remain the same except $\gamma_w = 0.5$, and this configuration is denoted as W95B. In the last case, all parameters are set to 1, $\lambda_w = \mu_w= \nu_w =\sigma_w =\delta_w = 1$, , and this configuration is denoted as the BT model \citep{Baggett1997}. It is important to note that these parameter sets significantly influence the system dynamics \citep{Baggett1997, Kalur2021}.

		The unconditional stability limit of the system can be calculated using equation (\ref{eq_Re_E}), which has the analytical solution:
		\begin{equation}\label{eq_W95_Re_E}
			Re_E = 2\sqrt{\lambda\,\mu}.
		\end{equation}
		\citep{Waleffe1995}. For the W95A and W95B models, $R_E = 20$ in the case of W95A and W95B model, , and for the BT model, $R_E = 2$. Below this critical value, the system is unconditionally stable, and the permissible perturbation level is infinite.
		
\begin{figure}
			\centering
			\subfigure[]{
				\includegraphics[scale=0.52]{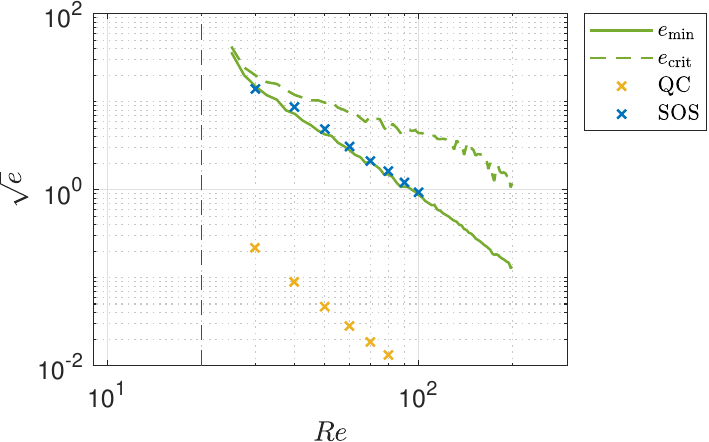}
				\label{fig:W95_g0p1_crit_energy}} 
			\hspace{5mm}
			\subfigure[]{
				\includegraphics[scale=0.52]{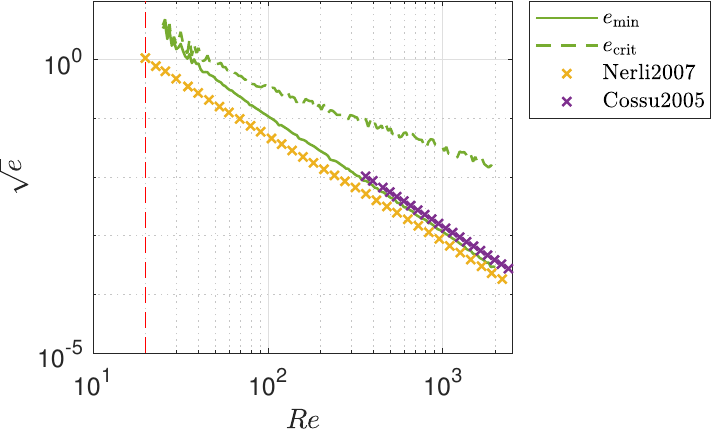}
				\label{fig:W95_g0p5_crit_energy}} 
			\hspace{5mm}
			\subfigure[]{
				\includegraphics[scale=0.52]{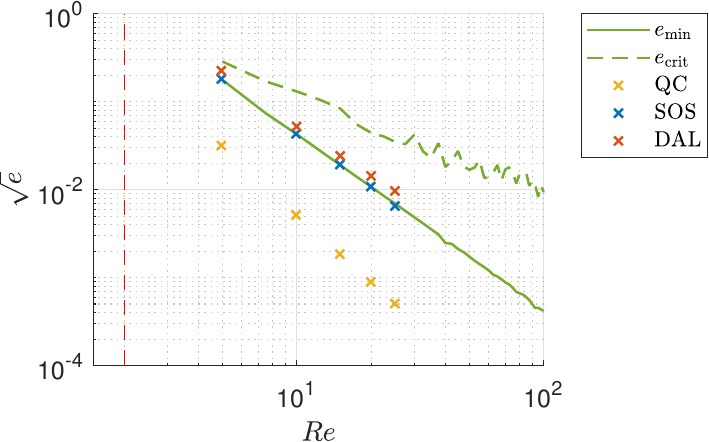}
				\label{fig:W95_BT_crit_energy}} 
			\caption{The square root of the smallest kinetic energy of the boundary of the region of attraction ($e_\mathrm{min}$) and the square root of the kinetic energy of the critical perturbation ($e_\mathrm{crit}$) as the function of Reynolds number in the case of W95A model (a), W95B model (b)  and BT model (c). The QC, SOS, DAL curves represent the results of \cite{Kalur2021}. The red, vertical dashed line represents the unconditional stability limit, $\Rey_E$. The best-fitting curves of $\sqrt{e_\mathrm{min}}$:  
				$\sqrt{e_\mathrm{min}}\approx 77102 \,\Rey^{-2.491}$ for W95A; $\sqrt{e_\mathrm{min}}\approx 1467.5 \,\Rey^{-2.043}$ for W95B ; $\sqrt{e_\mathrm{min}}\approx 4.2818 \,\Rey^{-2.0008}$ for BT model.}
			\label{fig:W95}
		\end{figure}

		The optimized transformation matrices are calculated for the W95A, W95B, and BT models over different ranges of Reynolds numbers: 25 to 200 for W95A model, 25 to 2000 for W95B model and 5 to 100 for BT model. For the W95A and BT models, the step size was set to 2.5, while for the W95B model, a logarithmic spacing was applied over 150 steps. Figure \ref{fig:W95} shows the largest inner radius of the region of attraction for the three models.

		The results are compared with other stability calculations methods. For the W95A and BT models, the proposed method yielded nearly the same permissible perturbation levels as the sum-of-squares (SOS) method used by \cite{Kalur2021}. 
		Furthermore, they applied the quadratic constraints (QC) method to the system, predicting significantly smaller regions due to the approximation of non-linear terms using bounds, although it required lower computational cost. 
		A comparative analysis of the accuracy of the QC method for the two-dimensional TTRD' model and these four-dimensional models suggests that the accuracy of the QC method deteriorates as the number of degrees of freedom of the model increases.
		In the case of W95B model, the result are  compared to the calculations of the generalized kinetic energy method by \citet{Nerli2007}. The presented novel implementation exhibited slight improvements due to the more general form of the energy function. Additionally, our results closely matched the non-linearly optimized minimal seeds calculated by \citet{Cossu2005}. (It is mentioned that \citet{Nerli2007} defined the kinetic energy with a multiplier of 1/2 which was compensated by a factor of $1/\sqrt{2}$ on the plots here.)  Similarly, both the SOS method and our result are very close to the optimized minimal seeds \citep{Kalur2021} of the BT model. In both cases, the close stability threshold energy and minimal seed energy values mean that the stability region is calculated within acceptable accuracy.
		
		At the same time, the solutions that are initialized outside the region of attraction tend to laminar state in the case of W95A model, which was also observed by \cite{Kalur2021}. This suggest that the true region of attraction is significantly larger than the predicted one. The larger region can be probably obtained utilizing higher-order energy (Lypunov) function. 
		
		For demonstration purposes of the method, four simulations are carried out using the BT parameters at $\Rey = 10$ initialized from values at the bound of the predicted region of attraction and values slightly outside of it. The optimal transformation matrix is given by 
		\begin{equation}
			\matr{S}_\mathrm{opt} \approx
			\begin{bmatrix} 0.748814 & -0.655534 & -0.0327291 & 0.00472172\\ -0.16196 & -0.0779668 & 0.00174343 & -0.00904493\\ 0.00406973 & -0.00425752 & 0.164989 & -0.00173407\\ -0.0485946 & 0.0213396 & 0.00172342 & 0.560516 \end{bmatrix} 
		\end{equation}
		and the corresponding critical vectors are 
		\begin{equation}
			\vect{q}_\mathrm{min}  \approx \begin{bmatrix}  0.00272966\\ 0.0373305\\ 0.0202733\\ 0.000371659\end{bmatrix}  \;\;\mathrm{and}\;\;%
			\vect{q}_\mathrm{crit} \approx \begin{bmatrix}  0.125899\\ 0.0144673\\ -0.0287946\\ -0.00806114  \end{bmatrix}.
		\end{equation}

		\begin{figure}
			\centering
			\subfigure[]{
				\includegraphics[scale=0.75]{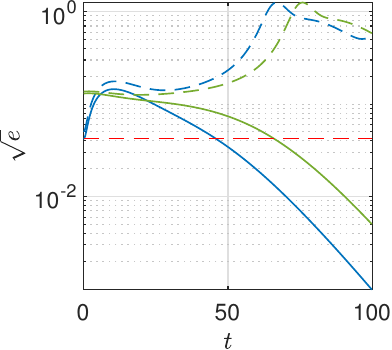}
				\label{fig:W95_BT_e_vs_time}} 
			\hspace{5mm}
			\subfigure[]{
				\includegraphics[scale=0.75]{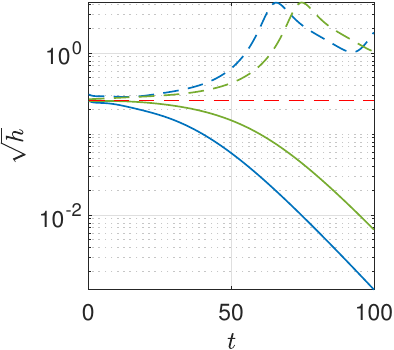}
				\label{fig:W95_BT_h_vs_time}} 
			\caption{The square root of the original kinetic energy (a) and the generalized kinetic energy (b) as the function of time. These solutions are obtained by BT model at $\Rey = 10$. The green continuous curve is initialized from $\vect{q}_\mathrm{min}$, the green dashed curve from $1.2 \vect{q}_\mathrm{min}$, the blue continuous curve from  $\vect{q}_\mathrm{crit}$, and the blue dashed curve from $1.05 \vect{q}_\mathrm{crit}$. The red, horizontal dashed line represents the permissible perturbation level.
			}
			\label{fig:W95_BT_eh_vs_time}
		\end{figure}
		
		Two solution are initialized with $\vect{q}_\mathrm{min}$ and $\vect{q}_\mathrm{crit}$, and the square root of their kinetic and generalized kinetic energy are plotted in figure \ref{fig:W95_BT_eh_vs_time} by blue and green colors, respectively.
		It can be observed that the generalized kinetic energy ($h_0$) is the same at the initial points, as both states are on the region of attraction hypersphere. Moreover, the generalized energy growth rates ($\mu_h$) are initially close to zero in both cases. However, this behavior is expected only in the case of a solution initialized by $\vect{q}_\mathrm{crit}$ following its definition. 
		As time progresses, both solutions exhibit a negative growth rate, tending towards the laminar equilibrium state. However, their initial original kinetic energies ($e_0$) differ due to the transformation of variables. Furthermore, a notable growth in kinetic energy ($\mu_e$) of the perturbation can be observed in the case of the solution initialized with $\vect{q}_\mathrm{min}$. Nevertheless, this classic energy eventually decays, as expected, since in another norm, its energy monotonically decreases over time.

		Two additional simulations were conducted, both initialized slightly outside of the predicted region of attraction:  $\vect{q}_0 = 1.2\,\vect{q}_\mathrm{min}$ and $\vect{q}_0 =1.05 \,\vect{q}_\mathrm{crit}$. It is noteworthy that in both cases, the solutions converge to a non-laminar equilibrium state. Specifically, the generalized kinetic energy experiences initial growth in both simulations, followed by oscillations around the non-laminar equilibrium state.
		It is important to observe that the kinetic energy in the simulation initialized by $\vect{q}_0 =1.2\,\vect{q}_\mathrm{min}$ grows significantly at the beginning due to non-normality. This growth leads to an energy level comparable to that of  $\vect{q}_\mathrm{crit}$. In contrast, in the other case, this pure non-modal growth is not observed. The original kinetic energy of the solution decays slightly at the beginning and increases only later.

		In summary, concerning the GKE results of the four-dimensional model, the predicted perturbation thresholds are validated as accurate in the cases of the BT and W95B models. However, it has been demonstrated to be overly conservative in the case of the W95A model. 

		\subsection{Poiseuille flow}
		\label{sec:Poi_flow}
		In the subsequent phase, a higher-order yet still low-dimensional model of the fluid dynamic system is developed to represent Poiseuille flow. This involves computing the Stokes eigenfunctions of a rectangular cuboid and determining the coefficients of the ordinary differential equation system using the Galerkin projection method. The Galerkin projection method, as established in previous research \citep{Nerli2006, Bergstrom1999}, proves to be an efficient approach for constructing low-order models. 
		
		The Stokes equations in non-dimensional form are given by:
		\begin{equation}\label{eq_Stokes}
			\pder{{u}_{i}}{t}=-\pder{{p}}{x_i}+\frac{1}{\Rey}\pder{^2 {u}_i}{x_j \partial x_j}
		\end{equation}
		and
		\begin{equation}\label{eq_cont}
			\pder{{u}_{i}}{x_i}=0
		\end{equation}
		where ${u}_{i}$ represents the non-dimensional velocity, $p$ is the non-dimensional pressure, and $x_i$ are the spatial coordinates: $x_1\in[0,L_x]; \, x_2\in[-1,1]; \,x_3\in[0,L_z]$, defining a rectangular cuboid.
		The eigenvectors can be obtained by assuming the following ansatz:
		\begin{equation}\label{eq_waveform}
			u_i = \hat{u}_{i}  \mathrm{e}^{ \lambda \,t }
		\end{equation}
		and solving the eigenvalue problem,
		\begin{equation}\label{eq_Stokes_eigen}
			\lambda\hat{u}_{i}=-\pder{\hat{p}}{x_i}+\frac{1}{\Rey}\pder{^2 \hat{u}_i}{x_j \partial x_j},
		\end{equation}
		for $\lambda$. The eigenvalues are negative real numbers expressing the dissipation rate of the mode. Furthermore, the eigenvectors are orthogonal, which proves advantageous for Galerkin projection.
		Given the linearity of the problem and assuming periodic solutions in $x_1$ and $x_2$ directions, solving the eigenvalue problem is conveniently achieved using complex Fourier series. The modes of the $\hat{u}_i$ velocity field can be expressed as follows:
		\begin{equation}\label{u_complex_wave}
			\tilde{u}_{i, j_m,k_m}(x_2)\mathrm{e}^{ \mathrm{i}(j_m \alpha_0 x_1 + k_m \beta_0 x_3)}
		\end{equation}
		where $\alpha_0 = 2 \upi / L_x$ and $\beta_0 = 2 \upi / L_z$ are the wavenumbers, and $j_m$, $k_m$ are the indices of the modes ranging from $-\infty$ to $\infty$.
		Substituting the complex wave form (\ref{u_complex_wave}) into the equations (\ref{eq_cont}) and (\ref{eq_Stokes_eigen}) leads to the following eigenvalue problem for each $j_m, k_m$ mode:
		\begin{equation} 
			\label{eq_Stokes_disc}
			\begin{bmatrix} L & 0 & 0 & -\mathrm{i}\alpha\\
				0 & L & 0 & -D_{x_2}\\
				0 & 0 & L & -\mathrm{i}\beta\\
				-\mathrm{i}\alpha & D_{x_2} & \mathrm{i}\beta & 0 \end{bmatrix} 
			\begin{bmatrix}\tilde{u}_1\\\tilde{u}_2\\\tilde{u}_3\\\tilde{p} \end{bmatrix} = 
			\lambda \begin{bmatrix} 1 & 0 & 0 & 0\\
				0 & 1 & 0 & 0\\
				0 & 0 & 1 & 0\\
				0 & 0 & 0 & 0
			\end{bmatrix} 
			\begin{bmatrix}\tilde{u}_1\\\tilde{u}_2\\\tilde{u}_3\\\tilde{p} \end{bmatrix} 
		\end{equation}
		where $\alpha = j_m\, \alpha_0$, $\beta = k_m\, \beta_0$ and $L = -(\alpha^2+\beta^2) +  D_{x_2}^2$ is the Laplace operator, where $D_{x_2}$ is the differential operator with respect to $x_2$. 
		
		The problem (\ref{eq_Stokes_disc}) can be discretized using the Chebyshev collocation method. The required boundary conditions involve stationary walls at the bottom and top of the domain, implying $\tilde{u}_i(\pm1) = 0$ for any $i$ velocity component. These conditions are enforced by removing the corresponding rows from the matrices. In this study, 100 Chebyshev collocation points are employed, a choice deemed accurate based on prior research \citep{Nagy2023}. 
		The discretized version of the equations (\ref{eq_Stokes_disc}) solved for the first $N_y$ modes with the largest $\lambda$ eigenvalues for $j_m \in [-N_x,N_x]$ and $k_m \in [-N_z,N_z]$ resulting in total $N_t = (2N_x+1)\, N_y\, (2N_z+1)$ number of modes. The calculation can be simplified, since in the case of complex conjugate wavenumber pairs ($j_{m, a} = -j_{m, b}$ and $k_{m, a} = -k_{m, b}$), the eigenvalues are the same and the eigenvectors are the complex conjugate of each other $\tilde{u}^{}_{i,j_m,k_m}=\tilde{u}^*_{i,-j_m,-k_m}$. The values of the parameters ($N_x, N_y,N_z$) vary across different models and will be provided later.
		Subsequently, the coefficients  $A_{i,j}$ and $Q_{i,j, k}$ are computed using the Galerkin projection method:
		\begin{equation}\label{eq_A_ij_Galorkin}
			A_{i_m,j_m} = \int_\varOmega \left(-{U}_{j}\pder{\hat{u}_{i, i_m}}{x_j}-\hat{u}_{j, i_m}\pder{{U}_{i}}{x_j}+\frac{1}{\Rey}\pder{^2 {u}_{i, i_m}}{x_j \partial x_j}\right) \hat{u}^*_{i, j_m}  \mathrm{d} \varOmega
		\end{equation} 
		\begin{equation}\label{eq_Q_ijk_Galorkin}
			Q_{i_m,j_m,k_m} = \int_\varOmega \left(-\hat{u}_{j, j_m}\pder{\hat{u}_{i, i_m}}{x_j}\right) \hat{u}^*_{i, k_m}  \mathrm{d} \varOmega
		\end{equation} 
		where $i_m,j_m$ and $k_m$ are the indices of the modes, ${U}_{i}$ denotes the velocity field of the base flow. For the Poiseuille flow investigated in this study, having only one non-zero velocity component:
		\begin{equation}\label{eq_U_Poi}
			U_1 = 1-x_2^2.
		\end{equation} 
		
		It is worth noting that
		\begin{equation}
			\int_\varOmega \pder{^2 {u}_{i, i_m}}{x_j \partial x_j}  \hat{u}^*_{i, j_m}  \mathrm{d} \varOmega = \lambda_{i_m}\, \delta_{i_m,j_m}
		\end{equation}  
		This is due to the fact that the velocity modes are solutions of the Stokes equation.

		The modes are substituted in the form (\ref{u_complex_wave}) and the integrals are evaluated utilizing Chebyshev collocation points. 
		%
		Since the eigenvectors are complex, the $A_{i,j}$ matrix and the $Q_{i,j,k}$ tensor are also complex. As a result, the previously derived gradients for the optimization procedure become invalid. However, this issue can be resolved by transforming the system into a real-valued one. Let $i_0$ represent the indices of the real-valued modes, $i_c$ the complex-valued modes, and $i_{cc}$ their corresponding complex conjugates. By rearranging the modes in the order $i_0, i_c, i_{cc}$, a transformation matrix $\matr{T}$  can be defined as follows:
		\begin{equation}\label{eq_complex_conj_transformation_matrix}
			\matr{T} = 
			\begin{bmatrix}
				T_{i_0, i_0}    & 	T_{i_0, i_c} & T_{i_0, i_{cc}}\\
				T_{i_c, i_0}    & 	T_{i_c, i_c} & T_{i_c, i_{cc}}\\		
				T_{i_{cc}, i_0} & 	T_{i_{cc}, i_c} & T_{i_{cc}, i_{cc}}\\
			\end{bmatrix}= 
			\begin{bmatrix}
				\matr{I}_{N_0\times N_0}    & 	\matr{0}_{N_0\times N_c} &	\matr{0}_{N_0\times N_c}\\
				\matr{0}_{N_c\times N_0}   & 	\frac{1}{\sqrt{2}}\matr{I}_{N_c\times N_c} & \frac{1}{\sqrt{2}}\matr{I}_{N_c\times N_c}\\		
				\matr{0}_{N_c\times N_0} &     \frac{1}{\mathrm{i}\sqrt{2}}\matr{I}_{N_c\times N_c} & -\frac{1}{\mathrm{i}\sqrt{2}}\matr{I}_{N_c\times N_c}\\
			\end{bmatrix}
		\end{equation} 
		Here, $N_0$ represents the number of real-valued modes, and $N_c$ represents the number of complex-valued modes (taking into account half of the complex-conjugate pairs). Applying the $\matr{S} = \matr{T}^{-1}$. Applying the $\matr{S}$ transformation matrix on the problem as described by equations (\ref{eq_A_tilde_def}) and (\ref{eq_Q_trans}) results in real-valued $A_{i,j}$ matrix and the $Q_{i,j,k}$ tensor, respectively. This transformation matrix can also be used to convert the transformed real coefficients back into the original complex coefficients of the complex-valued modes.

		\subsubsection{Results} 
		
		Two distinct configurations are explored in this study. In both cases, the dimensions of the domain are $L_x=2 \upi$ and $L_z = \upi$ , resulting in $\alpha_0 = 1$ and $\beta_0 = 2$. These domain sizes are chosen to ensure that the base wavenumbers ($\alpha_0,\, \beta_0$) are close to the critical values as determined by linear stability analysis ($\alpha = 1.02$)\citep{Orszag1971}	and standard non-linear stability analyis ($\beta = 2.04 $) \citep{Nagy2022}. Previous research by \citet{Reddy1998} also investigated Poiseuille flow on the same domain.
		In the first model, denoted as M1, the number of modes is set to $N_x = 1, N_y = 20, N_z = 1$ resulting in $N_t= 180$. In the second model, denoted as M2, the mode counts are $N_x = 1, N_y = 60, N_z = 1$, yielding $N_t= 540$. In these models, only the modification of the base flow is considered, while higher-order Fourier modes are neglected. It is important to note that these models may not capture the true behavior of the flow perfectly, but they serve as demonstrations of the GKE method on relatively high-order systems compared to previous studies. Increasing the number of modes significantly raises the computational cost due to the evaluation of non-linear terms, a well-known challenge in reduced-order models \citep{Sipp2020}. For the investigation of systems with more than 10,000 degrees of freedom, the current GKE method is not feasible.
		
		First, the linear stability limit ($\Rey_\mathrm{L}$) of the two models are determined, where the first eigenvalue of the linear part ($\matr{A}$) becomes positive. The influence of $N_y$ within the range of 10 to 100, as shown in  Table \ref{tab:Re_crit}. It is important to note that for fewer than 60 modes, the linear stability analysis is highly dependent on the number of modes due to the high sensitivity of the non-normal linear operator \citep{Trefethen2005} to numerical errors.
		Simultaneously, energy stability limit ($\Rey_\mathrm{E}$) of the system is less affected by the number of selected modes.
		
		These two limits are crucial for the model and can be relatively easily calculated. Below the energy stability limit, the flow is unconditionally stable, meaning that the radius of the region of attraction is infinite. On the other hand, beyond the linear stability limit, the flow is unconditionally unstable, and the radius of the region of attraction is 0. Between these two limits, the proposed method can be employed to calculate the conditional stability threshold.

		\begin{table}
			\begin{center}
				\def~{\hphantom{0}}
				\begin{tabular}{lcc}
					$N_y$  & $\Rey_\mathrm{L}$   &   $\Rey_\mathrm{E}$  \\[3pt]
					~10 & 1490080 & 49.8096 \\ 
					\textbf{~20} & \textbf{4544.82} & \textbf{49.6597} \\ 
					~30 & 2668.92 & 49.6306 \\ 
					~40 & 3452.81 & 49.6257 \\ 
					~50 & 4971.36 & 49.6228 \\ 
					\textbf{~60} & \textbf{5770.67} & \textbf{49.6220} \\ 
					~70 & 5995.41 & 49.6213 \\ 
					~80 & 5973.48 & 49.6210 \\ 
					~90 & 5947.59 & 49.6208 \\ 
					100 & 5930.30 & 49.6207 \\
				\end{tabular}
				\caption{The critical Reynolds number determined by linear stability analysis ($\Rey_L$) and energy stability analysis ($\Rey_E$) of the Poiseuille flow model with $L_x=2 \upi$ and $L_z = \upi$  and $N_x=1,N_z=1$. The values for the evaluated M1 and M2 models are indicated in bold.}
				\label{tab:Re_crit}
			\end{center}
		\end{table}
		
		\begin{figure}\centering
			\includegraphics[scale=0.75]{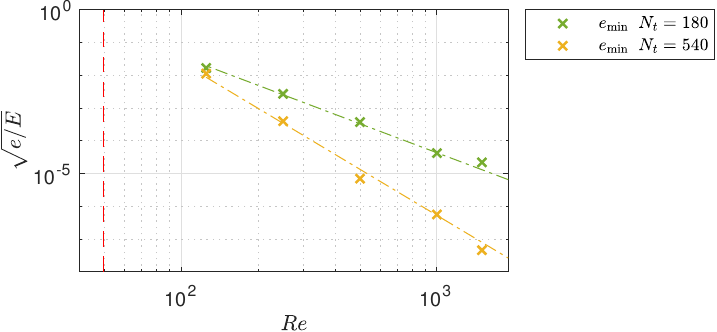}
			\caption{The square root of the ratio of allowable perturbation kinetic energy to the base flow kinetic energy as a function of Reynolds number for Poiseuille flow. The dimensions of the periodic domain are $2\upi \times 2 \times \upi$.}
			\label{fig:GALN_Poi}
		\end{figure}

		In the case of the previously defined models, M1 and M2, optimized transformation matrices are calculated for the following Reynolds numbers: 1500, 1000, 500, 250, and 125. To save computational time, the optimization process starts at the highest Reynolds number. Once the procedure converges, the next optimization at a lower Reynolds number is initialized with the previous optimal transformation matrix. It has been observed that if the procedure is initialized with a transformation matrix calculated at a lower Reynolds number, the generalized kinetic energy increases even for infinitesimally small amplitudes ($\gamma=0$). Modifying the initial matrix in this case would require additional computational cost. However, if an optimal transformation matrix from a higher Reynolds number is used, this issue does not arise.
		
		The results of optimization are plotted in Fig. \ref{fig:GALN_Poi}. The plot shows the square root of the allowable perturbation kinetic energy divided by the base flow kinetic energy ($E$), which is proportional to the ratio of perturbation velocity magnitude to the base flow velocity magnitude. This quantity is referred to here as the threshold amplitude ratio.
		While some previous studies aimed to find the minimal threshold energy or minimal seed for Poiseuille flow on systems with significantly higher degrees of freedom, a rough comparison between the results has been attempted. The amplitude ratio at different Reynolds numbers is presented in Table \ref{tab:thres_amp_scale}. Notably, the allowable perturbation amplitude ratio obtained in this study is significantly smaller than the threshold amplitude reported in previous studies.
		In the studies conducted by \cite{Lundbladh1994} and \cite{Reddy1998}, the base flow was perturbed with a prescribed or linearly optimized perturbation, and the threshold amplitude was investigated. However, non-linear optimization of the perturbation was not performed in these studies.
		\cite{Parente2022}, on the other hand, investigated the flow on a considerably larger domain and solved the non-linear minimal seed problem. They achieved a threshold amplitude one magnitude smaller at a slightly smaller Reynolds number compared to the results reported by \cite{Lundbladh1994} and \cite{Reddy1998}.
		Prior to comparing the results with the GKE method, it is crucial to acknowledge the differences in the approach: while previous studies focused on minimizing the necessary perturbation energy to induce transition, the current study maximizes the allowable perturbation.
		For the small system (M1), the amplitude ratio is only one magnitude smaller than the result of \cite{Parente2022}. However, for the larger system (M2), the values are three orders of magnitude smaller, indicating that the results obtained by the GKE method highly depend on the dynamical system's number of degrees of freedom. moreover, it should be emphasized that the dimensions of the systems in the cited papers were orders of magnitude larger.
		
		In the next step, power law functions are fitted to the threshold amplitude as a function of Reynolds number ($\sqrt{e}\propto\Rey^{\gamma_A}$) which has proven to be a good estimation in the case of Couette flow \citep{Duguet2013}. This approach has also been used in previously cited research. For Poiseuille flow, the exponents of the fitting are presented in Table \ref{tab:thres_amp_scale}, varying between -1.6 and -4.25 in studies \citep{Lundbladh1994, Reddy1998, Parente2022, Zhang2023}. Our predictions align closely with this established range, with exponents of -2.94 and -4.66 for the M1 and M2 models, respectively. The disparity between the exponents of M1 and M2 highlights that not only the amplitude but also the decay rate of the allowable perturbation amplitude decreases significantly as the number of unknowns increases.
		
		However, it's important to note that the cited models typically have significantly more degrees of freedom, and the fitted range of Reynolds numbers varies among the cited papers. Furthermore, due to the system's linear instability above a certain Reynolds number, the threshold amplitude as a function of Reynolds number must deviate from a simple power-law function.
		
		\begin{table}
			\begin{center}
				\def~{\hphantom{0}}
				\begin{tabular}{lcc}
					Source  & $\Rey=1000$   &  $\Rey=1500$  \\[3pt]
					\cite{Lundbladh1994}& - & 0.0053\\ 
					\cite{Reddy1998} & - & 0.00522\\
					\cite{Parente2022} &0.00144 & - \\
					M1, $N_t=180$ & 0.0000432 & 0.0000220\\
					M2, $N_t=540$ & $5.32\cdot 10^{-7}$  & $4.61\cdot 10^{-8}$\\
				\end{tabular}
				\caption{The threshold perturbation amplitude ratio ($\sqrt{e/E}$) for Poiseuille flow at different Reynolds numbers. Additional properties of the results can be found in Table \ref{tab:thres_amp}.}
				\label{tab:thres_amp}
			\end{center}
		\end{table}

		\begin{table}
			\begin{center}
				\def~{\hphantom{0}}
				\begin{tabular}{lcccc}
					Source  & Perturbation& Domain & $\Rey$ range& $\gamma_A$  \\[3pt]
					\cite{Lundbladh1994} & Oblique wave& $2\upi \times 2 \times 2\upi$&1500-5000& -1.75\\ 
					\cite{Reddy1998} & Oblique wave & $2\upi \times 2 \times 2\upi$&1500-5000& -1.6\\
					\cite{Parente2022} & Minimal seed & $250 \times 2 \times 125$&1000-1568& -4.25\\
					\cite{Zhang2023} & Minimal seed & $100 \times 2$ (2D)&2500-4500 & -3.8\\
					M1, $N_t=180$ & GKE stability & $2\upi \times 2 \times \upi$&125-1500& -2.94\\
					M2, $N_t=540$ & GKE stability & $2\upi \times 2 \times \upi$&125-1500& -4.66\\
				\end{tabular}
				\caption{Coefficients of the power law for the threshold amplitude ($\sqrt{e_\mathrm{min}}\propto\Rey^{-\gamma_A}$), half of the coefficient for the threshold energy.}
				\label{tab:thres_amp_scale}
			\end{center}
		\end{table}

		

		\section{Conclusion}
		\label{sec:Conclusion}
		
		In the study, an approach is introduced to establish the conditional stability limit of fluid flows by constructing a Lyapunov function. The core concept involves a linear transformation of the state variables and the definition of Generalized Kinetic Energy (GKE) as the inner product of these new variables. The method described here is analogous to the alteration of the inner product of the original state vectors, a modification explored by \citep{Nerli2007}. 
		
		The direct consequence of the transformation, the growth rate of generalized kinetic energy depends on the perturbation amplitude. This dependency enables us to calculate the threshold amplitude of stability, providing crucial insights into the system's behavior. Assuming an appropriate transformation matrix and a linearly stable system, we observe that the maximum potential growth rate of an infinitesimally small perturbation is negative. However, as the perturbation level increases, this growth rate steadily rises.
		
		The maximum potential growth of the system, in terms of perturbation level, can be classified into two distinct regions: initially, there is a constant phase characterized by a horizontal line, indicative of the dominance of linear dynamics at low perturbation levels. This phase is succeeded by a transitional region, leading to another straight line with a positive slope at higher perturbation levels, where the non-linear aspects of the system take precedence. The critical point occurs when the possible maximum growth rate of generalized kinetic energy intersects the zero line. This critical perturbation level signifies a threshold below which the flow remains stable, as the generalized kinetic energy diminishes, even though the standard kinetic energy may still increase.
		
		In the transformed state space, the attractive region is approximated as a hypersphere with a radius equal to the critical perturbation level. In the original state space, this region appears as a hyperellipsoid, with its smallest semiminor axis determining the maximum allowable perturbation kinetic energy. To optimize this perturbation kinetic energy level, the transformation matrix is fine-tuned. This optimization process involves deriving analytic gradients, rendering the method viable even for systems with a few thousand degrees of freedom.
		
		A crucial element in the calculations involves determining the global maximum of the potential growth rate among various perturbation states. To guarantee accuracy, the presented technique incorporates analytic gradients and the Hessian matrix, coupled with the use of multiple seed locations to ensure a comprehensive exploration of the solution space.
		
		The effectiveness of the method is demonstrated first on a relatively straightforward dynamical system: the turbulent flow's two-dimensional model, known as the TTRD' model, a simplified representation of turbulent flow. Here, the GKE approach adeptly approximates the region of attraction. Unstable solutions are identified with initial norms approximately 2\% larger than the predicted radius of the attraction region. Notably, the proposed method outperformed the quadratic constrained method, providing significantly more precise results.
		
		Moving forward, the GKE method is applied to three variations of the four-dimensional Waleffe model, each differing only in their parameters. In two instances, the presented GKE method predicted comparable allowable perturbation levels to those derived by \citet{Kalur2021} using the sum-of-squares method. The GKE approach outperformed the quadratic constrained method, producing results differing by orders of magnitude. Specifically, in the cases of W95B and BT parameter sets, unstable solutions \citep{Cossu2005, Kalur2021} are close to the predicted region of attraction, corroborating the accuracy of the GKE method.
		
		Finally, the method is extended to a reduced order model of the Poiseuille flow with 180 and 540 degrees of freedom. The predicted radius of the region of attraction decays similarly to the power law with the exponent of -2.94 and -4.66 in the small and large system, respectively. However, since the flow is linearly unstable above a certain Reynolds number, the decay of the radius must be faster the at higher Reynolds number.  

		In conclusion, the GKE method stands as a promising tool in the realm of fluid dynamics, providing accurate predictions for the conditional stability of linearly stable systems with a moderate number of degrees of freedom. While challenges persist in handling large systems, and further improvements of the method for flow modeling are necessary in the pursuit of understanding the conditional stability limits of fluid flows.

		
		\backsection[Acknowledgements]{The author is grateful to Yohann Duguet at CNRS for their helpful recommendations. }
		
		\backsection[Funding]{The research leading to these results received funding from the National Research Development and Innovation Office of Hungary under Grant Agreement no. K142675. }
		
		\backsection[Declaration of interests]{{\bf Declaration of Interests}. The author reports no conflict of interest.}
		
		
		\backsection[Author ORCIDs]{P. T. Nagy, https://orcid.org/0000-0002-8024-3824
		}
		
		
		\appendix
		
		\section{Numerical methods}\label{sec:app_numerical_methods}
		\subsection{The maximization of $\mu_{h}$}
		The critical aspect of the method lies in determining the maximum potential growth rate of generalized kinetic energy (\ref{eq_mu_h_max_def}).  In practical implementations, Matlab's \textit{fmincon} is employed, a tool that can significantly benefit from the provision of gradient and Hessian matrix of the cost function.
		The derivatives of the growth rate (\ref{eq_mu_h_calc2}) concerning the normalized state vector ($\tilde{r}_k$) are expressed as follows:
		\begin{align}
			\pder{\mu_h}{\tilde{r}_p}  &= 2\left(\tilde{A}_{p,i}+ \tilde{A}_{i,p}\right)\tilde{r}_i +\nonumber\\
			&+2\gamma\left(
			S_{i,j}^{-1} Q_{j,k,l} S_{k,p} S_{l,o}\tilde{r}_o \tilde{r}_i + 
			S_{i,j}^{-1} Q_{j,k,l} S_{k,m}\tilde{r}_m S_{l,p}\tilde{r}_i + 
			S_{p,j}^{-1} Q_{j,k,l} S_{k,m}\tilde{r}_m S_{l,o}\tilde{r}_o
			\right)
		\end{align}
		where $\tilde{A}_{i,j}$ is the transformed $A_{i,j}$ matrix defined in equation (\ref{eq_A_tilde_def}).
		Let us introduce the vectors $v_j = S^{-T}_{j,i}\tilde{r}_i$ and $\tilde{q}_i = S_{i,j}\tilde{r}_j$ to simplify the gradient:
		\begin{equation}
			\pder{\mu_h}{\tilde{r}_p}  = 2\left(\tilde{A}_{p,i}+ \tilde{A}_{i,p}\right)\tilde{r}_i 
			+2\gamma\left(
			S_{p,k}^{T} Q_{j,k,l} \tilde{q}_{l} v_j + 
			S_{p,l}^{T} Q_{j,k,l} \tilde{q}_{k} v_j +
			S_{p,j}^{-1} Q_{j,k,l} \tilde{q}_{k} \tilde{q}_l
			\right).
		\end{equation}
		The Hessian matrix of $\mu_h$ (\ref{eq_mu_h_calc2}) is given by:
		\begin{align}
			\pder{^2\mu_h}{\tilde{r}_p\partial \tilde{r}_q}  &= 2\left(\tilde{A}_{p,q}+ \tilde{A}_{q,p}\right) +\nonumber\\
			&2\gamma\left(
			S_{i,j}^{-1} Q_{j,k,l} S_{k,p} S_{l,q}\tilde{r}_i + 
			S_{q,j}^{-1} Q_{j,k,l} S_{k,p} S_{l,o}\tilde{r}_o +\right. \nonumber\\
			&S_{i,j}^{-1} Q_{j,k,l} S_{k,q} S_{l,p}\tilde{r}_i + 
			S_{q,j}^{-1} Q_{j,k,l} S_{k,m}\tilde{r}_m S_{l,p} \nonumber\\
			&\left. 
			S_{p,j}^{-1} Q_{j,k,l}  S_{k,q} S_{l,o}\tilde{r}_{o} + 
			S_{p,j}^{-1} Q_{j,k,l}  S_{k,m} \tilde{r}_{m} S_{l,q}
			\right)
			\label{eq:dmu_dr_Hessian}
		\end{align}
		By introducing the expressions:
		\begin{equation}
			B_{k,l} = Q_{j,k,l} v_j, \,\;\, 	C_{j,k} = Q_{j,k,l} \tilde{q}_l, \,\;\,  D_{j,l} = Q_{j,k,l} \tilde{q}_k,
		\end{equation}
		the equation (\ref{eq:dmu_dr_Hessian}) simplifies to:
		\begin{align}
			\pder{^2\mu_h}{\tilde{r}_p\partial \tilde{r}_q}  = 2\left(\tilde{A}_{p,q}+ \tilde{A}_{q,p}\right) +
			&2\gamma\left(
			S_{p,k}^{T} B_{k,l} S_{l,q} + 
			\left(S_{q,j}^{-1} C_{j,k} S_{k,p}\right)^T 
			+\right. \nonumber\\
			&\left(S_{q,k}^{T} B_{k,l} S_{l,p}\right)^T  + 
			\left(S_{p,j}^{-1} D_{j,l} S_{l,p}\right)^T \nonumber\\
			&\left. 
			S_{p,j}^{-1} C_{j,k}  S_{k,q} + 
			S_{p,j}^{-1} D_{j,l}  S_{l,q} 
			\right)
		\end{align}
		It's worth noticing that the Hessian matrix consists of the sum of four matrices and their transposes, resulting in a symmetric expression. This symmetry is expected due to the nature of second derivatives. From a practical perspective, only half of the expression needs to be calculated; the other half can be obtained by transposing the appropriate matrices.
		
		The optimization is constrained by the requirement that the transformed state vector should be unitary:
		\begin{equation}\label{eq_r_norm_constraint}
			c = \tilde{r}_i \tilde{r}_i - 1 = 0
		\end{equation}
		The gradient of the constraint is straightforward:
		\begin{equation}
			\pder{c}{\tilde{r}_i} = 2 \tilde{r}_i.
		\end{equation}
		The Hessian of the constraint (\ref{eq_r_norm_constraint}) is given by:
		\begin{equation}
			\pder{^2c}{\tilde{r}_i \partial \tilde{r}_j} = 2 \delta_{i,j}
		\end{equation}
		where $\delta_{i,j}$ is the Kronecker delta function and the right hand side is two times the identity matrix.
		
		\subsection{The maximization of $e_\mathrm{min}$}
		Maximizing $e_\mathrm{min}(S_{i,j})$ (\ref{eq:e_min_calc}) is a possibility, but it involves solving a complex, nonlinear equation system to calculate the gradients ($\mathrm{d}\, e_\mathrm{min}/ \mathrm{d}\, S_{i,j}$). An alternative approach is introducing the critical perturbation level $\gamma_\mathrm{crit}$ as as an additional variable of the cost function:  $e_\mathrm{min}(S_{i,j}, \gamma_\mathrm{crit})$ , constrained by the requirement that the maximum growth rate must be zero (\ref{eq_gamma_crit_def}).
		The gradients of kinetic energy for the allowable perturbation ($e_\mathrm{min}$) are given by:
		\begin{equation}
			\pder{e_\mathrm{min}}{S_{p,q}} = 2 \gamma_\mathrm{crit}^2 \tilde{r}_{\mathrm{min},q} S_pj \tilde{r}_{\mathrm{min},j} = 2 \gamma_\mathrm{crit}^2 \tilde{r}_{\mathrm{min},q} \tilde{q}_{\mathrm{min},p},
		\end{equation}
		and 
		\begin{equation}
			\pder{e_\mathrm{min}}{\gamma_\mathrm{crit}} = 2 \gamma_\mathrm{crit} \lambda_\mathrm{min}.	
		\end{equation}
		It's important to note that $\tilde{r}_{\mathrm{min},i}$ is a unit vector corresponds to the smallest eigenvalue of $S_{j,i} S_{j,k}$ matrix. This vector is distinct from $\tilde{r}_i$ used in subsequent expressions for calculating the maximum of $\mu_h$. $\tilde{r}_{\mathrm{min},i}$ depends solely on the transformation matrix.

		The derivatives of the constraint (\ref{eq_gamma_crit_def}) with respect of the elements of transformation matrix are
		\begin{align}\label{eq_dmu_dS}
			\pder{\mu_{h,\mathrm{max}}}{S_{p,q}} = 2\left(
			-\tilde{r}_i S_{i,p}^{-1}S_{q,j}^{-1} A_{j,m} S_{m,o}\tilde{r}_o + \tilde{r}_i S_{i,j}^{-1}A_{j,p} \tilde{r}_q
			\right)+\nonumber\\
			+2\gamma\left(
			-\tilde{r}_i S_{i,p}^{-1}S_{q,l}^{-1} Q_{l,m,o} S_{m,p} \tilde{r}_p S_{o,r} \tilde{r}_r +\tilde{r}_i S_{i,j}^{-1} Q_{j,p,l} \tilde{r}_q S_{l,o} \tilde{r}_o	
			\right. \nonumber\\
			\left.+ \tilde{r}_i S_{i,j}^{-1} Q_{j,k,p} S_{k,m} \tilde{r}_m \tilde{r}_q		
			\right)
		\end{align}
		where it is assumed that the inverse of the slightly perturbed transformation matrix can be approximated as:
		\begin{equation}
			\left(S_{i,j} + \delta S_{i,j} \right)^{-1} \approx S_{i,j}^{-1} - S_{i,k}^{-1}\,\delta S_{k,l}\, S_{l,j}^{-1}.
		\end{equation}
		The expressions (\ref{eq_dmu_dS2}) can be further simplified using the previously defined vectors and the transformed $A_{i,j}$ matrix:
		\begin{align}\label{eq_dmu_dS2}
			\pder{\mu_{h,\mathrm{max}}}{S_{p,q}} &= 2\left(
			-{v}_p \tilde{A}_{q,j}\tilde{r}_j + v_j A_{j,p} \tilde{r}_q
			\right)+\nonumber\\
			&+2\gamma\left(
			-{v}_p S_{q,l}^{-1} Q_{l,m,o} \tilde{q}_m \tilde{q}_o 
			+v_j  Q_{j,p,l} \tilde{r}_q \tilde{q}_l	
			+ v_j Q_{j,k,p} \tilde{q}_k \tilde{r}_q		
			\right)
		\end{align}
		Furthermore, the derivative of growth rate with respect to the perturbation level is
		\begin{equation}
			\pder{\mu_{h,\mathrm{max}}}{\gamma} =  2\, \gamma\, \tilde{Q}_{i,j,k}\,\tilde{r}_i\, \tilde{r}_j\,\tilde{r}_k , 
		\end{equation}
		which is simply the non-linear part of $\mu_h$, 
		\begin{equation}
			\pder{\mu_{h,\mathrm{max}}}{\gamma} =  \mu_{h,\mathrm{NL}}. 
		\end{equation}

		\bibliographystyle{jfm}
		\bibliography{biblography_V1}

	\end{document}